\documentclass[manuscript]{acmart} 
\makeatletter
\renewcommand\@formatdoi[1]{\ignorespaces}
\makeatother

\AtBeginDocument{%
  \providecommand\BibTeX{{%
    \normalfont B\kern-0.5em{\scshape i\kern-0.25em b}\kern-0.8em\TeX}}}

\setcopyright{iw3c2w3g}
\copyrightyear{2023}
\acmYear{2023}

\acmConference[CHI 2023]{Make sure to enter the correct
  conference title from your rights confirmation emai}{April 28,
  2023}{Hamburg, HH}
  
\acmBooktitle{Workshop WS32: Socially Assistive Robots as Decision Makers: Transparency, Motivations, and Intentions, ACM CHI Conference on Human Factors in Computing Systems ,
 April 28, 2023, Hamburg, HH} 
\usepackage[utf8]{inputenc}
\usepackage{wrapfig}

\begin{document}
\title[The Effects of Android Robots Displaying Emotion on Humans]{The Effects of Android Robots Displaying Emotion on Humans:
Interactions between Older Adults and Android Robots}

\author{Nora Hille}
\affiliation{%
  \institution{University of Siegen}
  \city{Siegen}
  \country{Germany}
}
\author{Berenike Bürvenich}
\affiliation{%
  \institution{University of Siegen}
  \city{Siegen}
  \country{Germany}
}
\author{Felix Carros}
\affiliation{%
  \institution{University of Siegen}
  \city{Siegen}
  \country{Germany}
}
\author{Mehrbod Manavi}
\affiliation{%
  \institution{University of Siegen}
  \city{Siegen}
  \country{Germany}
}
\author{Rainer Wieching}
\affiliation{%
  \institution{University of Siegen}
  \city{Siegen}
  \country{Germany}
\email{rainer.wieching@uni-siegen.de}
}
\author{Yoshio Matsumoto}
\affiliation{%
  \institution{National Institute of Advanced Industrial Science and Technology (AIST)}
  \city{Tokyo}
  \country{Japan}
\email{yoshio.matsumoto@aist.go.jp}
}

\author{Volker Wulf}
\affiliation{%
  \institution{University of Siegen}
  \city{Siegen}
  \country{Germany}
\email{volker.wulf@uni-siegen.de}
}

\renewcommand{\shortauthors}{Hille et al.}

\date{February 2023}

\begin{abstract}

%

Often robots are seen as a means to an end to fulfill a logical objective task. Android robots, on the other hand, provide new possibilities to fulfill emotional tasks and could therefore be integrated into assistive scenarios. We explored this possibility by letting older adults and stakeholders have a conversation with an android robot capable of expressing emotion through facial expressions. The study was carried out with a wizard-of-oz approach and data collected with a mixed methods approach.  
We found that the participants were encouraged to speak more with the robot due to its smile. Simultaneously, many ethical questions were raised about transparency and manipulation. Our research can give valuable insight into the reaction of older adults to android robots that show emotions. 

\end{abstract}




\maketitle

\section{Introduction \& Related Research
}
Human-Robot interaction has made great advances in the last decade, with social robots being put to practice in different social situations (e.g., care settings \cite{carros2020exploring,10.1145/3491102.3517463, helmsocial}, religion \cite{trovato2021religion, trovato2019communicating, loffler2021blessing}, rehabilitation \cite{langer2021emerging, kellmeyer2018social}. Still, emotions are seen as one of the aspects that separate humans from machines.
Emotions are integral to developing empathy and understanding the intent of the interaction partner. Sadness, joy, and fear are used by humans to determine that the creatures they interact with are deserving of respect and care \cite{torre2020if}. Still, we also developed fine-tuned abilities to notice signs of disingenuousness \cite{reed2018face}.
Next to tone and gestures, we mainly use facial expressions to detect and infer the emotions of our human counterparts \cite{crivelli2018facial}.
A phenomenon called emotional contagion describes the fact that emotions can even be transmitted to another person \cite{hatfield2014new}. This often includes facial mimicry, the copying of a facial expression that was displayed by a counterpart \cite{rymarczyk2019empathy}. Primary emotions are joy, sadness, anger, fear, disgust, and surprise. They often describe a spontaneous and intuitive reaction toward an event and have specific facial expressions associated with them across cultures \cite{ekman1992argument}. 

There is insightful research shedding light on the beneficial effect of displayed emotions in human-robot interaction. For example, Chuah and Yu could show that the display of joy and surprise by service robots positively influenced potential customers \cite{chuah2021future}. As it is rather complicated to artificially create realistic facial expressions, many researchers explored different channels to convey emotions \cite{haring2011creation}. Haring et al. tried communicating emotions through sounds, body movements, and even eye color using a Nao robot but concluded that the latter was ineffective, as it was too far removed from the way emotions are naturally displayed \cite{haring2011creation}. Torre et al. found that not only visual expressions of happiness like smiles increased trust, but also conversational agents that spoke with a tone implying a smile were met with a higher level of trust \cite{torre2020if}. Likewise, Beck et al. could show that even a Nao robot that lacks the ability to alter its face could communicate emotions using its body language \cite{beck2010towards}.
Another way to create realistic emotions is through using animated avatars \cite{noel2009interpreting}. These avatars can be portrayed on a display or projected into a three-dimensional space \cite{matovelle2018interaction}.
There are also robots that try to very closely mimic natural human faces. These so-called androids can mimic facial expressions very closely as they have controllable joints in their faces \cite{ishiguro2007android}. Nishio et al. even built a robot that exactly resembled one of the researchers to directly compare the effect this robot would have on people to the one the real person has \cite{nishio2007geminoid}.
In order to shed light on how older adults react to the display of emotion by a human-like robot, we let people interact with a very realistic android. Through interviews and surveys, we captured the impression these basic emotions caused and their effects on trust.

\section{Methods}
In the study, 12 participants interacted with an android robot. The robot looked like a young woman and it was able to display four primary emotions – joy, sadness, anger and surprise. 
The model of the android was an A-Lab Android Standard Model AL-G109ST-F.


\begin{wrapfigure}{R}{0.35\textwidth}
\includegraphics[width=1\linewidth]{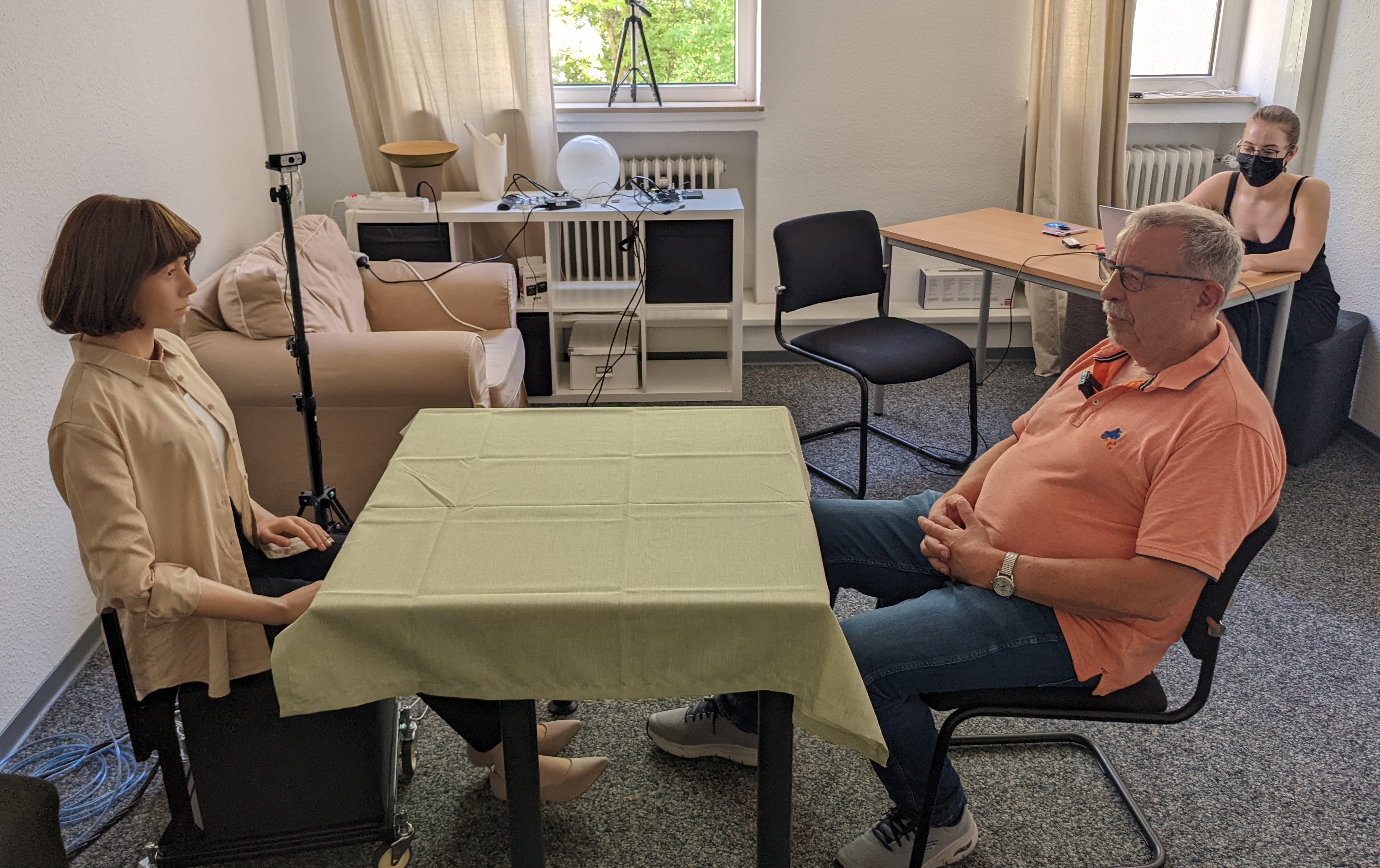} 
\caption{Android Robot \& participant}
\label{fig:andoroid}
\end{wrapfigure}

A living lab situation was created in which participants had the opportunity to interact with the android for about thirty minutes. The study was designed as a Wizard-of-Oz interaction, meaning that the android’s phrases were controlled and written by the researchers \cite{dahlback1993wizard, carros2022ethical}. The facial expression was a mix of the Wizard-of-Oz simulating emotions and an automatic mode making micro movements such as simulating breathing or looking around the room. During the interaction, the android was seated across from the participant. Meanwhile, two researchers sat behind them out of view, and a conversation was started through a series of loosely defined questions and statements. The topics started with casual small talk like “What is your favorite season?” to more personal questions like “What are you looking forward to today?” to very deep questions like “What is your greatest fear?”.
Eight participants who were all above the age of 60 were invited as well as four stakeholders, whose ages ranged from 48 to 67, who could offer a professional opinion on the effect a robot displaying emotion could have on older adults. In the following, these stakeholder participants will be denoted with SH and regular participants with P.  
Directly after the interactions, semi-structured interviews were conducted on how the interaction was perceived and how the participants would evaluate potential future uses, especially in regard to care homes. Interviews lasted 30-60 minutes and were later transcribed and deductively coded. Using a reflexive thematic analysis \cite{braun2021can} initial categories were formulated and inductively expanded while reviewing the interviews. The codes were also discussed among researchers. As main categories, it was looked at the way the participants rated the speech, gestures, and facial expressions of the android. Making a distinction of whether an aspect was seen as positive or negative in each one. It was also coded how often the participants mentioned an emotion. Further main categories were potential problems, visions for the future, and the usage of the android within a care home – each of which was further divided into subcategories.
\section{Results}
Despite the fact that the android was capable of displaying the emotions of happiness, surprise, anger and sadness, most participants only recalled that the robot showed joy. All the participants who interacted with the android in the university mentioned its smile and so did two of the stakeholders.
SH1 stated that the android’s smile encouraged her to keep talking. It was also described as \textit{“pretty cute, only just alluded but visible.”} (P8) and \textit{“nice”} (P4) and P3 explained that she thought the smile was astute and that it brought her joy. P7 was more tentative and described that the android had tried to smile, implying he would not consider it a success.
Sadness and surprise were only recalled by P1, P3 and P7. This disparity can be explained by the fact that the facial expressions were only used when they were appropriate in the context of the conversation, which was true more often for happiness than it was for either sadness or surprise.
SH2 stressed that the important part about the facial expression was not that they were present, but rather that they were displayed in the right moment within the context of the conversation.
SH4 saw the facial expressions as providing an additional path of communication.
SH1 emphasized the fact that the android smiling at her made her realize that she understood what the topic was about. She also mentioned that this encouraged her to talk more. P8 mentioned that she considered the nodding of the android to have a reassuring effect on her and stressed that this aspect, in particular, made the android appear very human-like. Reaction was one of the most important keywords in regard to emotion. Whenever they spoke positively about the display of emotions, the participants stressed how the android, which they often even called by its human name, reacted with a smile or mirrored their own emotions (P1, P2).
P2, P3 and P8 stated that they were positively surprised by the fact that the android was capable of displaying facial expressions. P6 said that the expressions made the robot likable and human-like.

When asked how the emotions of the robot could be expanded, the participants mentioned that they could imagine a robot displaying embarrassment (P3), compassion (P4), laughing out loud (P3, SH2), or crying (SH1).
P6 stressed that it was important that the robot never showed aggression though and always stayed gentle. P8 and SH2 also emphasized that the robot’s main objective should be to be calm and not upset anyone interacting with it. They also wished the android would show more intrinsic curiosity (P4) by asking questions and inquiring about them.
In contrast to this, P1, P2 and P8 all wished for the robot to have its own opinion and be able to discuss controversial topics and even explained that they would want the robot to disagree, set boundaries and actively urge the counterpart to discuss with them. P8 even elaborated that she would consider it problematic for humans to constantly interact with a counterpart that always agreed with them and never opposed them.
SH2 imagined a specific future scenario in which the robot could help older adults overcome traumatic losses in their lives.
P8, on the other hand, doubted that the robots could ever do certain emotional tasks. Her examples were hugging a person in need or helping a patient with dementia to return to the real world by reacting with empathy and understanding.

A prominent concern is that emotional display will eventually blur the lines between humans and robots to a point where it is no longer easy to tell them apart. 
SH3 explained that people in care homes were at an especially great risk to make that mistake due to their health situation. P3 saw it as important to ensure that robots and humans could always be told apart.
P8 and SH3 also mentioned that they would think it was very likely for an older adult to fall in love with a robot. P6 stated \textit{“when you are with a robot there will somehow always be relationships that develop. And whether they can bring fulfillment or lead to disappointment we cannot tell.”}
SH1 also said: \textit{“A robot will not be able to react right when the other person starts to cry or show emotions.”}
SH2 did not see any risks in humans mistaking robots for others humans.
Although concerns were also raised by SH1 she did draw the conclusion that it would ultimately be the “lesser evil” to have some older adults mistake a robot for a human, if they could at least be connected to an actual human by the robot. 
SH1 stated that she was sure many people would accuse a robot that could display realistic gestures, facial expressions and emotions of lying to the users and deceiving them.
P6 and P8 both agreed that they would not see any threat of manipulation in the current state of the robot. P8 and SH3 also elaborated that she considered it non-problematic to create the illusion of emotion as long as the users had a positive experience. P7 added that many books and movies also created illusions and that those were not seen as ethically questionable.

\section{Discussion \& Conclusion}
As emotions are expressions of internal states and robots do not have consciousness, it is controversial to make them display emotions \cite{giger2019humanization, coeckelbergh2011emotional, ojha2018essence}. 
It could be argued that even humans use facial expressions as tools and communicational shortcuts \cite{crivelli2018facial}, so letting robots utilize them will only positively affect the ability to interact with them. Furthermore, many researchers are of the opinion that human-robot interaction can be facilitated by letting robots show emotion \cite{Azeem2012emotions,loffer2018multimodel}, and the feedback of the participants suggests the same. Still, we believe  it is essential always to make the human partner aware that it is interacting with a robot. 
In the context of using androids in care homes, it is necessary to remember that people with impairments could be at risk of mistaking an android for a human. This can lead to expectations that are ultimately not met.
SH1 and SH3 raised an important point with their assertion that even if the use of robots as emotional support surely is ethically difficult, we have to ensure to weigh the potential threats against the benefits. A robot counterpart could be an enjoyable distraction that is welcomed by care home residents and might contribute to a decrease in loneliness. 
It can be argued that robots might have some advantages of providing emotional support to humans as they are equipped with certain “superpowers” like endless patience or unconditional subordination that humans do not possess \cite{ welge2016better}. Still, it is necessary that new laws and guidelines accompany the potential use of robots in care homes that ensure enough staff supervises the interactions and that the artificiality of the robot is openly communicated to everyone who interacts with it \cite{carros2022care, schwaninger2022video}.

The smile of the robot was generally seen as beneficial. It leads the participants to rate the android as more human and likable. This is most likely not because the participants thought the robot was actually happy but caused by the fact that smiling at someone signals understanding. This type of encouragement can extend the motivation to engage in conversation, which is why participants also said it increased their willingness to speak to the android. This is consistent with the theory that smiles are not predominantly used to convey happiness but more often simply signify affiliate intent \cite{fang2019unmasking}.
When it comes to other and potentially more negative emotions, there are a lot of risks and benefits to be weighed. On the one hand, there is a desire for humans to interact with a real counterpart that is capable of discussion and offering dissenting views, one that would have reason to display emotions like anger and disgust. On the other hand, there is the valid assertion that robots should only display emotions that will evoke a positive reaction in the interactant.
Likewise, the display of sadness or pity comes with the threat of leading to expectations for the relationship that will ultimately be disappointed or create a situation that is so emotionally charged that the robot cannot help the human to overcome them.

This study is limited by the number of participants and the lengths of the interactions. For now, it has to be clearly stated, though, that any of the participants experienced no difficulty in telling that they were not, in fact, interacting with a human, indicating that the robot is not that human-like after all \cite{carros2023not}. This might change in the future; therefore, it might be reasonable to establish internationally binding guidelines to differentiate robots and humans visually. In addition, the display of negative emotions, such as sadness, fear, etc., needs further investigation.
\bibliographystyle{Bib/ACM-Reference-Format}
\bibliography{References}

\end{document}